\documentclass{article}
\usepackage{spconf}
\usepackage{diagbox}
\usepackage{amssymb}
\usepackage[T1]{fontenc}
\usepackage{amsmath, graphicx, hyperref, multirow, caption, booktabs, array, makecell, bm}
\usepackage{subcaption}
\usepackage[hyphenbreaks]{breakurl}
\usepackage{adjustbox}
\usepackage{enumitem}
\title{MDCTCodec: A Lightweight MDCT-based Neural Audio Codec towards High Sampling Rate and Low Bitrate Scenarios}
\name{Xiao-Hang Jiang, Yang Ai\sthanks{Corresponding author. This work was funded by the National Nature Science Foundation of China under Grant 62301521, the Anhui Provincial Natural Science Foundation under Grant 2308085QF200, and the Fundamental Research Funds for the Central Universities under Grant WK2100000033.}, Rui-Chen Zheng, Hui-Peng Du, Ye-Xin Lu, Zhen-Hua Ling}
\address{National Engineering Research Center of Speech and Language Information Processing, \\University of Science and Technology of China, Hefei, P. R. China \\{\small \tt \ \{jiang\_xiaohang, zhengruichen, redmist, yxlu0102\}@mail.ustc.edu.cn}, \\ {\small \tt \{yangai, zhling\}@ustc.edu.cn}}

\begin{document}
\ninept
\maketitle
\begin{abstract}

In this paper, we propose MDCTCodec, an efficient lightweight end-to-end neural audio codec based on the modified discrete cosine transform (MDCT). 
The encoder takes the MDCT spectrum of audio as input, encoding it into a  continuous latent code which is then discretized by a residual vector quantizer (RVQ). 
Subsequently, the decoder decodes the MDCT spectrum from the quantized  latent code and reconstructs audio via inverse MDCT. 
During the training phase, a novel multi-resolution MDCT-based discriminator (MR-MDCTD) is adopted to discriminate the natural or decoded MDCT spectrum for adversarial training. 
Experimental results confirm that, in scenarios with high sampling rates and low bitrates, the MDCTCodec exhibited high decoded audio quality, improved training and generation efficiency, and compact model size compared to baseline codecs. 
Specifically, the MDCTCodec achieved a ViSQOL score of 4.18 at a sampling rate of 48 kHz and a bitrate of 6 kbps on the public VCTK corpus. 
\end{abstract}
\vspace{-1mm}
\section{Introduction}
\vspace{-1mm}
\label{sec:intro}
Audio codec, a vital component in digital audio processing, serves the dual purpose of encoding and decoding audio signals. 
It plays a critical role in reducing the data size required to represent audio while maintaining acceptable decoded audio quality. 
Audio codecs can be widely applied in audio communication \cite{salami1994toll}, audio compression \cite{brandenburg1994iso} and several downstream tasks such as speech synthesis \cite{wang2023neural,zhang2023speechtokenizer,borsos2023audiolm,mehta2024matcha,kim2021conditional}. 
Lately, there has been a growing focus on high-sampling-rate codecs, which are important for high-quality communications or speech synthesis. 
However, the increase in sampling rates inevitably leads to a high bitrates and negatively impacts on compression efficiency and transmission costs.

Traditional audio codecs, e.g., Opus \cite{valin2013high} and EVS \cite{dietz2015overview}, combine traditional coding tools, such as LPC \cite{o1988linear} and CELP \cite{schroeder1985code}, to ensure highly efficient audio coding. 
However, when challenged by high sampling rates and low bitrates, traditional codecs have not shown good performance due to insufficient bits, resulting in the appearance of audible artifacts \cite{disch2016intelligent}.

Recently, with the rapid development of deep learning, an increasing number of neural audio codecs have been proposed. 
Early neural audio codecs, such as Lyra \cite{kleijn2021generative}, encode compressed representations with traditional codecs first, then reconstruct the decoded waveform through a neural vocoder. 
They can run at lower bitrates, but their quality is not satisfactory.

Recently, research on end-to-end neural audio codecs has gained more popularity.
As for end-to-end neural audio codecs, the encoder, quantizer and decoder are all designed based on neural networks and trained jointly. 
SoundStream \cite{zeghidour2021soundstream} pioneers this type of codecs by using an encoder to directly encode the waveform, employing residual vector quantizer (RVQ) \cite{vasuki2006review} for discretization, and finally using a decoder to reconstruct the decoded waveform. 
Encodec \cite{defossez2022high} and HiFi-Codec \cite{yang2023hifi} are improved based on the framework of SoundStream, using more advanced training and quantization strategies.
However, these codecs are mostly designed for low sampling rates (e.g., 16 kHz and 24 kHz) and can hardly deal with 48 kHz audio which possesses exceptionally high temporal resolution. 
Thus, the above shortcomings limit their applications at high sampling rates.
DAC \cite{kumar2024high} can perform coding on 44.1 kHz audio, but it requires a bitrate of 8 kbps. 
Additionally, in our preliminary experiments, the generation efficiency of DAC was also very slow. 
Although AudioDec \cite{wu2023audiodec} is capable of operating at 48 kHz, the bitrate remains high (12.8 kbps) and it requires the assistance of a neural vocoder, resulting in a complicated training process. 
Additionally, these codecs often require hundreds of downsampling/upsampling operations due to their direct manipulation of waveforms, leading to high model complexity and training difficulties.

Hence, an alternate end-to-end parametric neural audio codec has been proposed to discretize audio spectral features rather than the raw audio waveforms. 
The windowed spectra significantly reduce the temporal resolutions compared to the original waveforms, allowing these codecs to achieve low bitrate compression with simple downsampling/upsampling operations with fewer model parameters. 
For example, APCodec \cite{ai2024apcodec} operates on the short-time Fourier transform (STFT) results. Compared to the STFT spectrum, the modified discrete cosine transform (MDCT) spectrum is simpler and has higher compression efficiency. 
For instance, MDCTNet \cite{davidson2023high} employs the MDCT spectrum as its coding object. 
Both APCodec and MDCTNet are designed for a 48 kHz sampling rate. 
But MDCTNet \cite{davidson2023high} adopts a complex recurrent structure and does not use neural quantization strategies, resulting in high bitrates (20-32 kbps). 
Although APCodec \cite{ai2024apcodec} supports low bitrate (6 kbps) compression, it requires parallel amplitude and phase streams, which increases model complexity and constrains efficiency.

In view of the simplicity and high compressibility of the MDCT spectrum, we propose MDCTCodec, which uses the MDCT spectrum as the coding object and finally reconstructs the audio waveform via inverse MDCT (IMDCT). 
However, it distinguishes itself from MDCTNet \cite{davidson2023high} by its ability to accomplish low-bitrate compression at high sampling rates. 
The encoder and decoder of the MDCTCodec both use a modified ConvNeXt v2 network as the backbone, linked by an RVQ. 
Furthermore, we also innovatively propose a multi-resolution MDCT-based discriminator (MR-MDCTD) for the adversarial training of MDCTCodec. 
Experiments verify that MDCTCodec achieves high decoded audio quality, extremely fast training and generation efficiency, and fewer model parameters (i.e., lightweight model), compared to baseline neural audio codecs. 
At a sampling rate of 48 kHz and a bitrate of 6 kbps, our proposed MDCTCodec achieves a ViSQOL score \cite{chinen2020visqol} of 4.18 on the public VCTK corpus \cite{veaux2017superseded}, and it can achieve 123$\times$ and 16.9$\times$ real-time generation speed on GPU and CPU, respectively.

This paper is organized as follows.
In Section \ref{sec:format}, we briefly review the waveform-coding-based neural audio codecs and spectral-coding-based neural audio codecs, respectively. 
In Section \ref{sec: propose}, we provide details of the model structure and training criteria of the proposed MDCTCodec.
In Section \ref{sec: Experiments}, we present our experimental results.
Finally, we give conclusions in Section \ref{sec: Conclusion}.
\vspace{-1mm}
\vspace{-1mm}
\vspace{-1mm}
\section{Related Works}
\label{sec:format}
\vspace{-1mm}

\subsection{Waveform-Coding-based Neural Audio Codec} 
\vspace{-1mm}
Early neural audio codecs mostly use waveforms as the coding object. 
SoundStream \cite{zeghidour2021soundstream} is one of the earliest end-to-end waveform-coding-based neural audio codecs. 
Compared to traditional codecs, it significantly reduces the bitrate while maintaining high decoded audio quality. 
SoundStream consists of an encoder, a quantizer, and a decoder. 
It uses SEANet \cite{tagliasacchi2020seanet} as the backbone network for both the encoder and decoder to directly encode and decode the waveform, while the quantizer employs an RVQ. 
In terms of training, SoundStream employs a adversarial training strategy, utilizing two types of discriminators, i.e., a waveform-based discriminator and an STFT-based discriminator. 
For the specific coding process of SoundStream, the waveform inputted into the encoder goes through multiple downsampling convolutional layers, resulting in a low sampling rate continuous codes. 
The compressed continuous codes are quantized into discrete codes using an RVQ, and then decoded into the decoded waveform by a decoder whose structure is symmetric to that of the encoder with same-ratio upsampling. 

Subsequent improvements to waveform-coding-based neural audio codecs have mostly revolved around the SoundStream framework. 
For example, Encodec \cite{defossez2022high} improves the training loss, while HiFi-Codec \cite{yang2023hifi} improves the quantization strategy by proposing group RVQ (GRVQ) to alleviate codebook redundancy and reduce bitrate. 
DAC \cite{kumar2024high}, on the other hand, improves RVQ to enhance codebook utilization and applies it to high-sampling-rate (44.1 kHz) audio coding tracks. 
AudioDec \cite{wu2023audiodec} employs a two-stage training mode and leverages a vocoder to achieve high-sampling-rate (48 kHz) audio coding. 

However, due to the high temporal resolution of waveforms, the waveform-coding-based neural audio codecs require hundreds of times of downsampling and upsampling operations. 
This is particularly unfriendly for high-sampling-rate waveforms, leading to high model complexity and increased training difficulty. 
Additionally, directly modeling the waveform also significantly impacts generation efficiency.


\subsection{Spectral-Coding-based Neural Audio Codec} 

To overcome the aforementioned issues, spectral-coding-based neural audio codecs have recently emerged, with APCodec \cite{ai2024apcodec} being a representation. 
Since audio amplitude and phase spectra can be converted from and to the audio waveform via STFT and ISTFT, APCodec uses amplitude and phase spectra as the coding objects instead of the waveform. 
This allows it to achieve audio coding at a 48 kHz sampling rate and 6 kbps bitrate. 
Both the encoder and decoder of APCodec have parallel structures that separately process amplitude and phase spectra, while the quantizer also uses RVQ. 
To accurately decode the phase, APCodec employs a phase parallel estimation architecture and phase anti-wrapping loss proposed in \cite{ai2023neural}. 
In APCodec, the STFT frame shift for extracting amplitude and phase spectra is set to 40 samples. 
Therefore, it only requires 8$\times$ downsampling/upsampling operations to achieve a 320$\times$ compression of the waveform, which is consistent with the settings of most waveform-coding-based neural audio codecs. 
In terms of model training, APCodec adopts an adversarial training strategy, employing both multi-period discriminator (MPD) \cite{kong2020hifi} and multi-resolution discriminator (MRD) \cite{siuzdak2023vocos}. 

Despite achieving impressive compression performance, the APCodec still has the following issues. 
In terms of the model structure, APCodec has to encode and decode both amplitude and phase in parallel, which requires a more complex dual-path network structure, resulting in double model complexity. 
Regarding the training criteria, APCodec uses two discriminators and multiple loss terms, which inevitably affect training efficiency. 
Our proposed MDCTCodec uses the MDCT spectrum as the coding target, employing only a single network and a single discriminator with a simple loss function, which can effectively improve training/generation efficiency and reduce model complexity.


\begin{figure}
  \centering
  \includegraphics[width=0.95\linewidth]{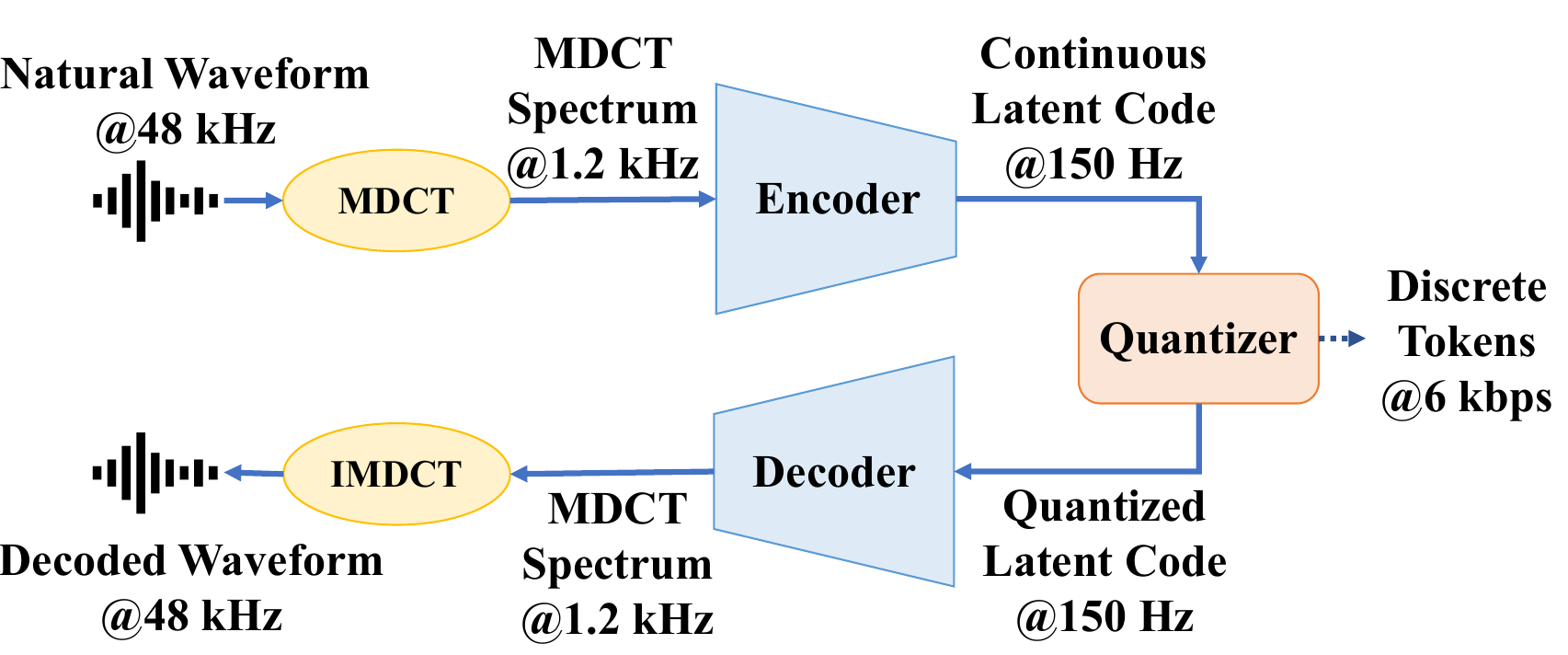}
  \caption{An overview of the proposed MDCTCodec.}
  \label{fig: Overview}
\end{figure}
\vspace{-1mm}

\begin{figure*}[htp]
  \centering
  \includegraphics[width=0.95\linewidth]{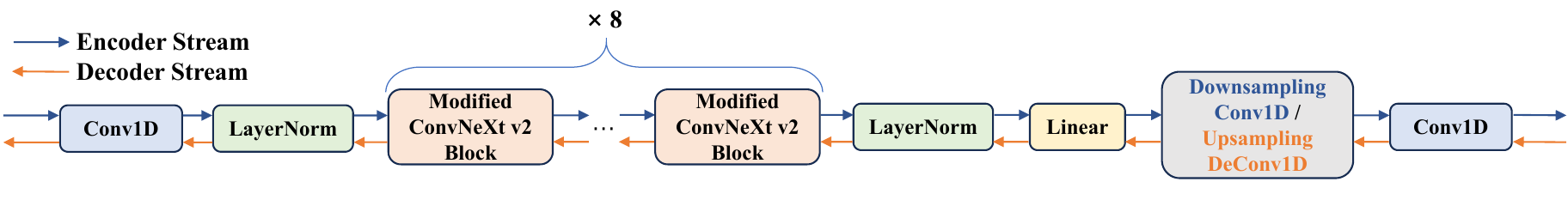}
  \caption{Details of structures of the encoder and decoder in the MDCTCodec. Here, \emph{Conv1D} and \emph{DeConv1D} represent 1D convolutional layer and 1D deconvolutional layer, respectively.}
  \label{fig: encoderdecoder}
\end{figure*}
\vspace{-1mm}
\vspace{-1mm}
\section{Proposed Method}
\label{sec: propose}
\vspace{-1mm}
\vspace{-1mm}
\subsection{Overview}
Figure \ref{fig: Overview} shows an overview of the proposed MDCTCodec model. 
At the encoder side, the input natural waveform $\bm{x}=[x_1,x_2,\dots,x_T]^\top \in \mathbb{R}^{T}$ with a high sampling rate of $S$ is first converted to an MDCT spectrum $\bm{X}\in\mathbb{R}^{N\times K}$ via MDCT.
The frame length, frame shift and frequency bin number of the MDCT operation are $W_L$, $W_S$ and $K$, respectively. 
Here, $T$ is the number of the waveform samples, and $N=T/W_S$ is the number of frames in the MDCT spectrum. 
In the implementation of MDCT, the following conditions must be satisfied: $W_S=K=W_L/2$. 
For $n$-th windowed waveform $\bm{x}_n=[x_{(n-1)K+1},\dots,x_{(n+1)K}]^\top \in \mathbb{R}^{2K}$, the MDCT spectrum value at the $n$-th frame and $k$-th frequency bin is as follows: 
\begin{align}
  \bm{X}[n,k] &= \sum_{l=1}^{2K} \bm{x}_n[l]\cos\left[\frac{\pi}{K}\left(l-\frac{1}{2}+\frac{K}{2}\right)\left(k-\frac{1}{2}\right)\right],
\end{align}
where $n=1,2,\dots,N$ and $k=1,2,\dots,K$.
Then, the encoder encodes the MDCT spectrum $\bm{X}\in\mathbb{R}^{N\times K}$ into a continuous code $\bm{C}\in\mathbb{R}^{(N/R)\times K'}$ with lower temporal resolution (i.e., $R>1$) and lower dimension (i.e., $K'<K$). 
$R$ represents the downsampling rate of the encoder and the upsampling rate of the decoder. 
An RVQ consisting of $Q$ vector quantizers (VQs) with a codebook size of $M$, is used to discretize $\bm{C}\in\mathbb{R}^{(N/R)\times K'}$, generating discrete tokens with a bitrate of $\frac{S}{W_s\cdot R}\cdot Q \cdot \log_2M$ (bps). 
These tokens are then converted into quantized code $\hat{\bm{C}}\in\mathbb{R}^{(N/R)\times K'}$ for decoding by querying the codebooks.  
Finally, the decoder decodes the MDCT spectrum $\hat{\bm{X}}\in\mathbb{R}^{N\times K}$ from $\hat{\bm{C}}$, and recovers the decoded waveform $\hat{\bm{x}} \in \mathbb{R}^{T}$ via IMDCT. 
For the $n$-th windowed waveform $\hat{\bm{x}}\in\mathbb{R}^{2K}$, the $l$-th sample value is calculated as follows. 
\begin{align}
  \hat{\bm{x}}_n[l] &= \frac{1}{K}\sum_{k=1}^{K} \hat{\bm{X}}[n,k]\cos\left[\frac{\pi}{K}\left(l-\frac{1}{2}+\frac{K}{2}\right)\left(k-\frac{1}{2}\right)\right],
\end{align}
where $l=1,2,\dots,2K$ and $n=1,2,\dots,N$. 
$\hat{\bm{X}}[n,k]$ represents the predicted MDCT spectrum value at the $n$-th frame and $k$-th frequency bin. 
Finally, the overlap-add algorithm is applied to $\hat{\bm{x}}_1,\hat{\bm{x}}_2,\dots,\hat{\bm{x}}_N$ to construct the final decoded waveform $\hat{\bm{x}}$.

\subsection{Model Structure}

Figure \ref{fig: encoderdecoder} illustrates the detailed architecture of the encoder and decoder, both of which are designed based on convolutional frameworks. 
A modified ConvNeXt v2 network \cite{ai2024apcodec} is employed as the backbone, which consists of 8 cascaded modified ConvNeXt v2 blocks. 
This block is modified from an original block in the field of image processing \cite{woo2023convnext} and has been validated to be suitable for audio coding tasks \cite{ai2024apcodec}. 
It employs a residual connection structure, with the core modules including 1D depth-wise convolutional layer, layer normalization operations, linear layer, global response normalization (GRN) \cite{woo2023convnext} layer and Gaussian error linear unit (GELU) activation \cite{hendrycks2016gaussian}.
As shown in Figure \ref{fig: encoderdecoder}, 1D convolutional layers, linear layers, and layer normalization operations are employed at the input/output ends of the network for feature preprocessing/post-processing. 
At the end of the encoder, a 1D convolutional layer further downsamples the input features by a factor of $D$, and finally, another 1D convolutional layer reduces the output dimensionality to $K'$. 
Except for the downsampling convolutional layer being replaced by the upsampling deconvolutional layer, the decoder and encoder are basically mirror symmetric. 
Therefore, the MDCTCodec employs a single non-parallel structure, which can effectively reduce the number of model parameters and improve computational efficiency compared to APCodec \cite{ai2024apcodec}.

\subsection{Training Criteria}

\begin{figure}
  \centering
  \includegraphics[width=0.95\linewidth]{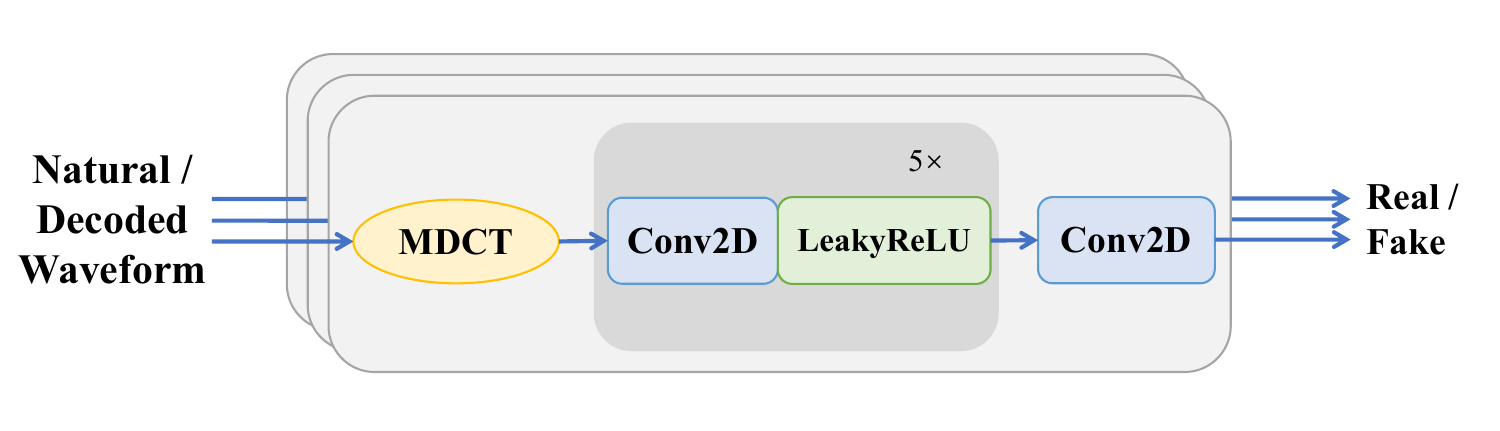}
  \caption{Details of the structure of the MR-MDCTD. Here, \emph{Conv2D} represents 2D convolutional layer.}
  \label{fig: MR-MDCTD}
\end{figure}

We adopt a generative adversarial network (GAN) \cite{goodfellow2014generative} based training strategy for MDCTCodec. 
A novel MR-MDCTD is proposed to engage in adversarial training with the generator (i.e., the encoder, quantizer and decoder in Figure \ref{fig: Overview}). 
Figure \ref{fig: MR-MDCTD} illustrates the structure of the MR-MDCTD, which is composed of three parallel sub-discriminators. 

\vspace{-1mm}

The $i$-th sub-discriminator $D_i$ first extracts the MDCT spectrum from input $\bm{x}$ or $\hat{\bm{x}}$ through a pre-configured MDCT whose frame length, frame shift and frequency bin number are $W_{Li}$, $W_{Si}$ and $K_i$, respectively. 
The MDCT spectrum is initially processed by three cascaded blocks composed of 2D convolutional layers and leaky rectified linear unit (LeakyReLU) \cite{maas2013rectifier} activations. Finally, a 2D convolutional layer with a single channel is employed to produce the discriminative score. 
These sub-discriminators utilize different MDCT configurations, enabling MR-MDCTD to perform discrimination on MDCT spectra across multiple resolutions. 
The adversarial loss in hinge form is employed.
For the generator and discriminator, the adversarial losses are defined as follows: 
\begin{align}
\label{equ: GAN_G}
\mathcal L_{\rm adv-G} = \sum_{i=1}^3 \mathbb{E}_{\hat{\bm{x}}}[\max(0,1-D_i(\hat{\bm{x}}))],
\end{align}
\begin{align}
\label{equ: GAN_D}
\mathcal L_{\rm adv-D} = \sum_{i=1}^3 \mathbb{E}_{{\bm{x}},\hat{\bm{x}}}[\max(0,1-D_i(\bm{x}))+\max(0,1+D_i(\hat{\bm{x}}))].
\end{align}
Furthermore, the feature matching loss $\mathcal L_{\rm FM}$ \cite{kumar2019melgan} is also applied.

\begin{table*}
\centering
    \caption{Experimental results on decoded audio quality evaluations for compared codes at three bitrates on the VCTK test set. The \textbf{bold} and \underline{underline} numbers indicate optimal and sub-optimal results, respectively.}
    \resizebox{\textwidth}{!}{
    \begin{tabular}{c | c c c | c c c | c c c}
        \hline
        \hline
        & \multicolumn{3}{c|}{Bitrate = 6 kbps} & \multicolumn{3}{c|}{Bitrate = 9 kbps} & \multicolumn{3}{c}{Bitrate = 12 kbps}\\
        \cline{2-10}
         & LSD (dB) $\downarrow$ & STOI $\uparrow$ & ViSQOL $\uparrow$ & LSD (dB) $\downarrow$ & STOI $\uparrow$ & ViSQOL $\uparrow$ & LSD (dB) $\downarrow$ & STOI $\uparrow$ & ViSQOL $\uparrow$ \\
         \hline
         SoundStream & 0.937 & 0.794 & 3.11 & 0.917 & 0.830 & 3.26 & 0.916 & 0.831 & 3.32\\
         Encodec & 1.039 & 0.793 & 3.31 & 0.892 & 0.846 & 3.48 & 0.885 & 0.860 & 3.51\\
         HiFi-Codec & 0.913 & 0.816 & 3.33 & / & / & / & 0.829 & \underline{0.941} & 3.88 \\
         AudioDec & 0.847 & 0.811 & 3.98 & 0.839 & 0.815 & 4.13 & 0.831 & 0.825 & 4.14 \\
         DAC & 0.841 & \textbf{0.906} & 3.81 & 0.826 & \textbf{0.936} & 3.95 & 0.815 & \textbf{0.954} & 4.06 \\
         APCodec & \textbf{0.818} & 0.875 & \underline{4.07} & \textbf{0.810} & 0.889 & \underline{4.14} & \textbf{0.796} & 0.901 & \underline{4.26} \\
         MDCTCodec & \underline{0.825} & \underline{0.891} & \textbf{4.18} & \underline{0.816} & \underline{0.923} & \textbf{4.26} & \underline{0.812} & 0.934 & \textbf{4.32} \\
        \hline
        \hline
    \end{tabular}}
\label{tab: Objective evaluation results}
\end{table*}

\begin{table}
  \caption{Experimental results on efficiency and complexity evaluations for compared codecs at 6 kbps on the VCTK test set. Here, \emph{s/e} represents second per epoch. The \textbf{bold} and \underline{underline} numbers indicate optimal and sub-optimal results, respectively.}
  \label{tab: efficiency}
  \centering
  \begin{tabular}{ c  c c c c }
    \toprule
    \multirow{2}{*}{}  & \multicolumn{1}{c}{RTF} & \multicolumn{1}{c}{RTF} & \multicolumn{1}{c}{Training} &  \multirow{1}{*}{Model}  \\& (GPU) $\downarrow$ & (CPU) $\downarrow$ & Time $\downarrow$& Size $\downarrow$ \\
    \midrule
    SoundStream           &0.0148&0.231&2567 s/e &83.2M     \\
    Encodec               &0.0149&0.232&1980 s/e &83.2M       \\
    HiFi-Codec            &0.0178&0.878&5741 s/e &243M \\
    AudioDec              &0.0132&0.771& 1343 s/e &108M        \\
    DAC                   &0.0195&2.47& 3471 s/e &282M        \\
    APCodec               &\underline{0.0112}&\underline{0.173}& \underline{752 s/e} &\underline{65.4M}    \\
    MDCTCodec             &\textbf{0.0081}&\textbf{0.059}& \textbf{ 140 s/e} &\textbf{26.2M}    \\
    \bottomrule
  \end{tabular}
\end{table}

In addition to the GAN-based loss, we also incorporate spectral-level losses and quantization losses. 
The spectral losses include 1) MDCT spectrum loss $\mathcal L_{\rm MDCT}$, defined as the mean squared error (MSE) between the decoded MDCT spectrum $\hat{\bm{X}}$ and the natural one $\bm{X}$, i.e.,
\begin{align}
\label{equ: MDCT Loss}
\mathcal L_{\rm MDCT}=\dfrac{1}{NK}\mathbb{E}_{\left(\hat{\bm{X}},\bm{X}\right)}||\hat{\bm{X}}-\bm{X}||_F^2,
\end{align}
where $||\cdot||_F$ denotes the Frobenius norm; 
2) mel spectrogram loss $\mathcal L_{\rm Mel}$, which is the sum of the MSE and mean absolute error (MAE) between the mel spectrogram $\hat{\bm{M}}\in\mathbb R^{N\times K_M}$ extracted from $\hat{\bm{x}}$ and the mel spectrogram $\bm{M}\in\mathbb R^{N\times K_M}$ extracted from $\bm{x}$, i.e., 
\begin{align}
\label{equ: M Loss}
\mathcal L_{\rm Mel}=\dfrac{1}{NK_M}\cdot\mathbb{E}_{\left(\hat{\bm{M}},\bm{M}\right)}\left( \left\lVert \hat{\bm{M}}-\bm{M}\right\rVert_1 + \left\lVert \hat{\bm{M}}-\bm{M}\right\rVert_F^2\right),
\end{align}
where $K_M$ is the dimensionality of the mel spectrogram and $||\cdot||_1$ denotes the L1 norm. 
Besides, we also apply the quantization loss, consisting of the codebook loss $\mathcal L_{\rm cb}$ and the commitment loss $\mathcal L_{\rm com}$ borrowed from \cite{yang2023hifi} to the RVQ, aiming to narrow the quantization error. 

The overall generator loss for jointly training the encoder, quantizer and decoder is defined as follows.
\begin{equation}
    \begin{aligned}
   \mathcal L_{\rm G} &=  \mathcal L_{\rm adv-G} + \mathcal L_{\rm FM} + \lambda_{\rm MDCT} \mathcal L_{\rm MDCT} + \\ & \lambda_{\rm Mel} \mathcal L_{\rm Mel} + \lambda_{\rm cb} \mathcal L_{\rm cb} + \lambda_{\rm com} \mathcal L_{\rm com},
    \end{aligned}
\end{equation}
where $\lambda_{\rm MDCT}$, $\lambda_{\rm Mel}$, $\lambda_{\rm cb}$ and $\lambda_{\rm com}$ represent hyperparameters. 
During the training process, the generator (i.e., the encoder, quantizer and decoder) and discriminator (i.e., the MR-MDCTD) are trained alternately using $\mathcal L_{\rm G}$ and $\mathcal L_{\rm adv-D}$.

\vspace{-1mm}
\section{Experiments}
\vspace{-1mm}
\label{sec: Experiments}
\subsection{Experimental Setting}
\label{subsec: setting}
Following the experiment data settings of AudioDec \cite{wu2023audiodec} and APCodec \cite{ai2024apcodec}, we also utilized the VCTK dataset \cite{veaux2017superseded} in our experiment\footnote{Audio samples of the proposed MDCTCodec can be accessed at \href{https://pb20000090.github.io/MDCTCodecSLT2024/}{https://pb20000090.github.io/MDCTCodecSLT2024/}.}. 
The sampling rate of the data is 48 kHz, and the duration is approximately 43 hours. 
40,936 utterances from 100 speakers and 2,937 utterances from the remaining 8 unseen speakers were used for training and testing. 
The configuration settings for the MDCT operation in the MDCTCodec model are $W_L=80$, and $W_L=K=40$. 
A cosine window was used for framing. 
In Figure \ref{fig: encoderdecoder}, for both the encoder and decoder, except for the last module, the output feature dimensions of all other modules were 256. 
The intermediate dimensions for modified ConvNeXt v2 blocks were 512. 
The output dimensions of the last 1D convolutional layer in the encoder and decoder are $K'=32$ and $K=40$, respectively. 
All convolutional layers had a kernel size of 7. 
For the MR-MDCTD, the configuration for the three sets of MDCT is as follows: $W_{L1}=400$, $W_{S1}=K_1=200$; $W_{L2}=100$, $W_{S2}=K_2=50$; $W_{L3}=40$, $W_{S3}=K_3=20$. 
The channel sizes for the first 5 2D convolutional layers were all set to 64, and their kernel sizes were 7$\times$5, 5$\times$3, 5$\times$3, 3$\times$3 and 3$\times$3, respectively.  
The last output 2D convolutional layer had 1 channel, and its kernel size was 3$\times$3. 
The hyperparameters for loss functions were set as $\lambda_{MDCT}=250$, $\lambda_{Mel}=45$, $\lambda_{cb}=10$ and $\lambda_{com}=0.25$. 
The training was conducted using the AdamW optimizer \cite{loshchilov2018decoupled}, with $\beta_1 = 0.8$ and $\beta_2 = 0.99$. 
The learning rate decayed by a factor of 0.999 after each epoch, starting from an initial learning rate of 0.0002. 
The batch size was 48 and each training step utilized a truncated waveform length of 7,960 samples. 
The number of overall training iterations is 200k and all experiments were conducted on a single NVIDIA RTX 3090 GPU. 
\vspace{-1mm}
\subsection{Evaluation Metrics}
\vspace{-1mm}
Both objective and subjective metrics are used for evaluating the performance of compared codecs from different perspectives. 

We employed three objective metrics to evaluate the decoded audio quality. We utilized commonly used log-spectral distance (LSD), short-time objective intelligibility (STOI), and visual speech quality listener tool (ViSQOL) \cite{chinen2020visqol}, which respectively focused on the amplitude spectrum quality, intelligibility, and overall audio quality. 
Additionally, to evaluate the generation and training efficiency, the real-time factor (RTF) and training time were separately measured. 
The RTF stands for the ratio of generation time to audio's actual duration. 
The training time is defined as the time taken to complete one epoch of training. 
To measure the model complexity, we also compiled the model size of codecs. 
Model complexity was indicative of computational requirements and potential storage constraints, thereby providing an important perspective on the practicality and scalability. 
Therefore, a lightweight model is favored. 

To assess the subjective quality, we also conducted ABX preference tests on the Amazon Mechanical Turk platform\footnote{\href{https://www.mturk.com}{https://www.mturk.com.}} to compare our proposed MDCTCodec with other baseline codecs one by one.  
In each ABX test, at least 30 native English-speaking listeners evaluated 20 paired utterances, which were randomly selected from the test sets decoded by the codecs being compared. 
The ABX experiment required listeners to judge which of the utterances in each pair had better speech quality, or if there was no preference. 
In addition to calculating the average preference scores, we also used the $p$ value of a $t$-test to measure the significance of the difference between two compared codecs.


\vspace{-1mm}
\begin{figure}[t] 
\centering
\captionsetup[subfigure]{labelformat=empty}
\begin{subfigure}{0.22\textwidth}    
    \centering
    \caption{Natural waveform}
    \includegraphics[width=\linewidth]{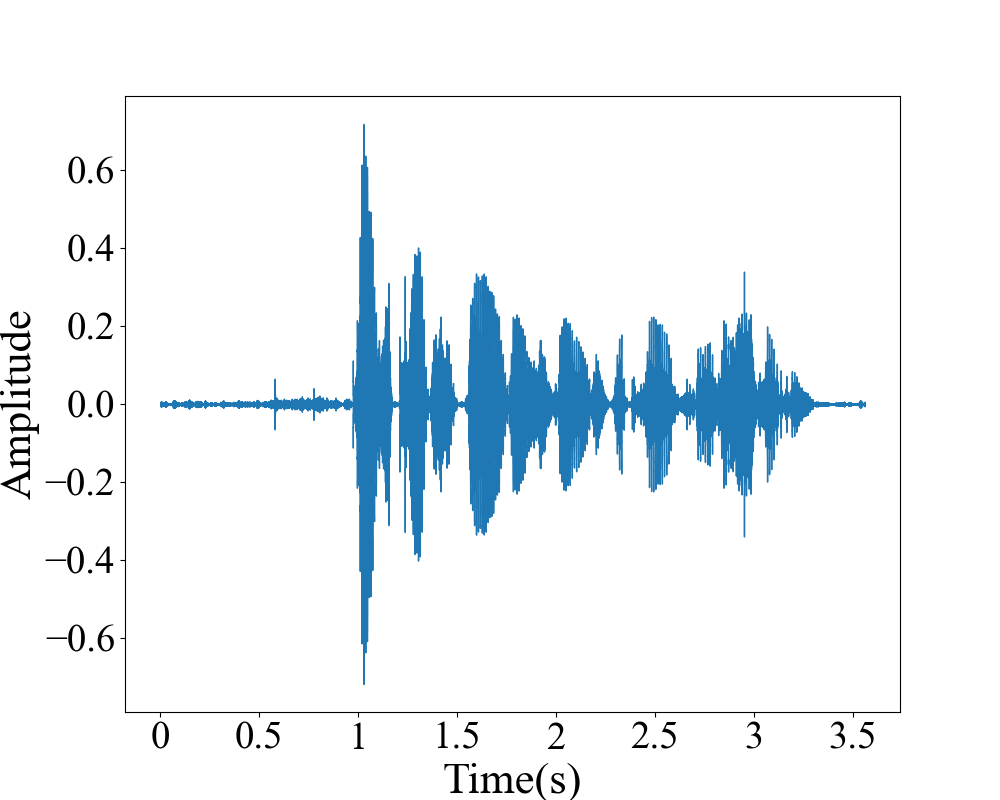}
    \label{fig:sub1}
\end{subfigure}\hfill
\begin{subfigure}{0.22\textwidth}
    \centering
    \caption{Decoded waveform}
    \includegraphics[width=\linewidth]{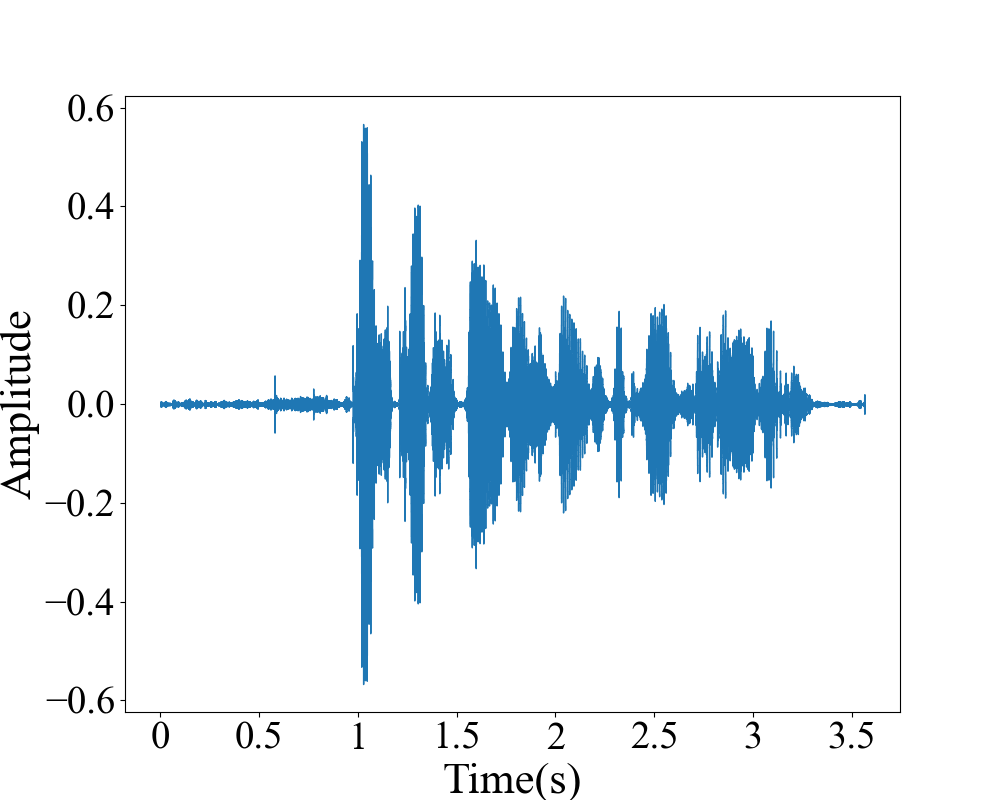}
    \label{fig:sub2}
\end{subfigure}
\begin{subfigure}{0.22\textwidth}
    \centering
    \caption{Natural MDCT spectrum}
    \includegraphics[width=\linewidth]{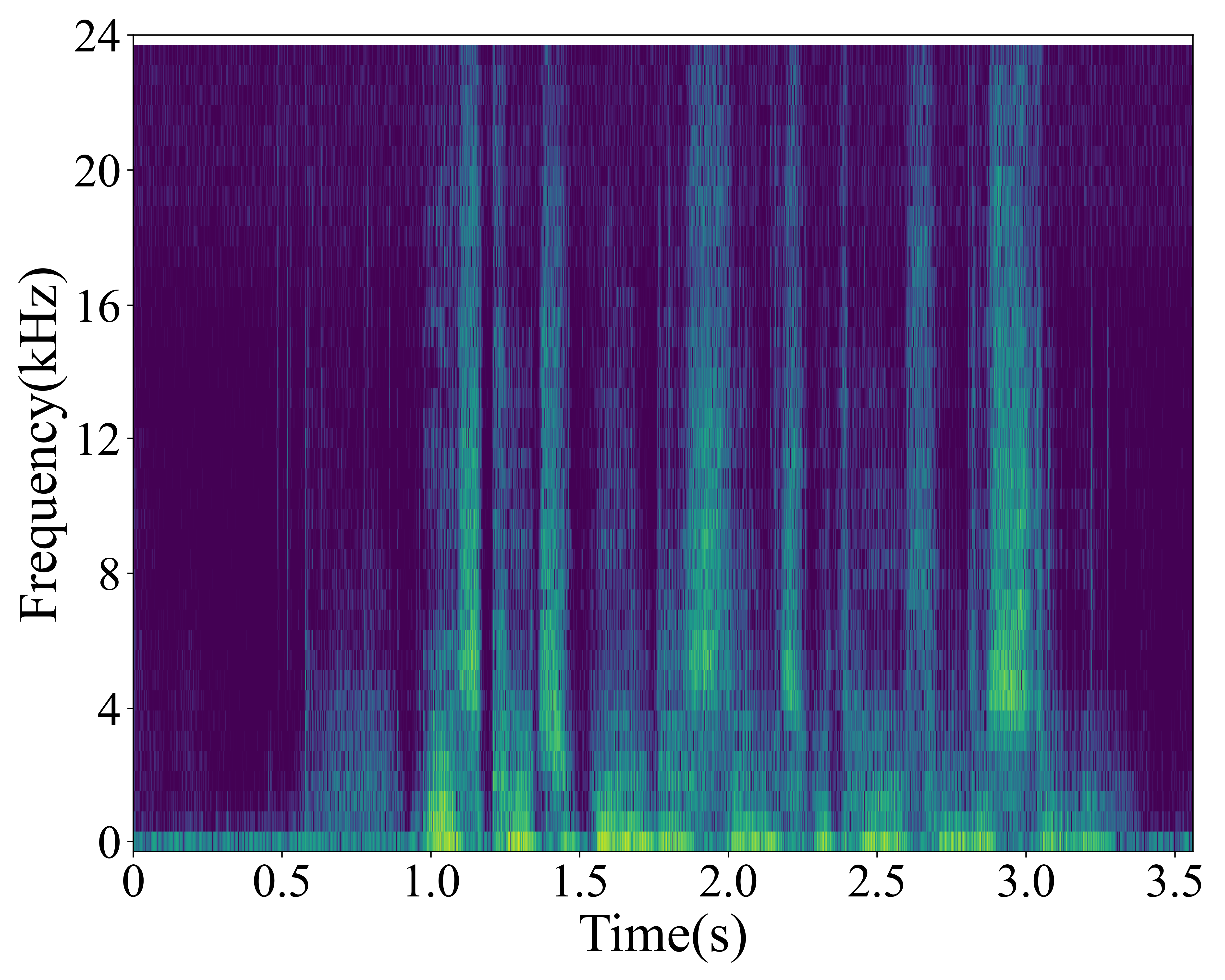}
    \label{fig:sub3}
\end{subfigure}\hfill
\begin{subfigure}{0.22\textwidth}
    \centering
    \caption{Decoded MDCT spectrum}
    \includegraphics[width=\linewidth]{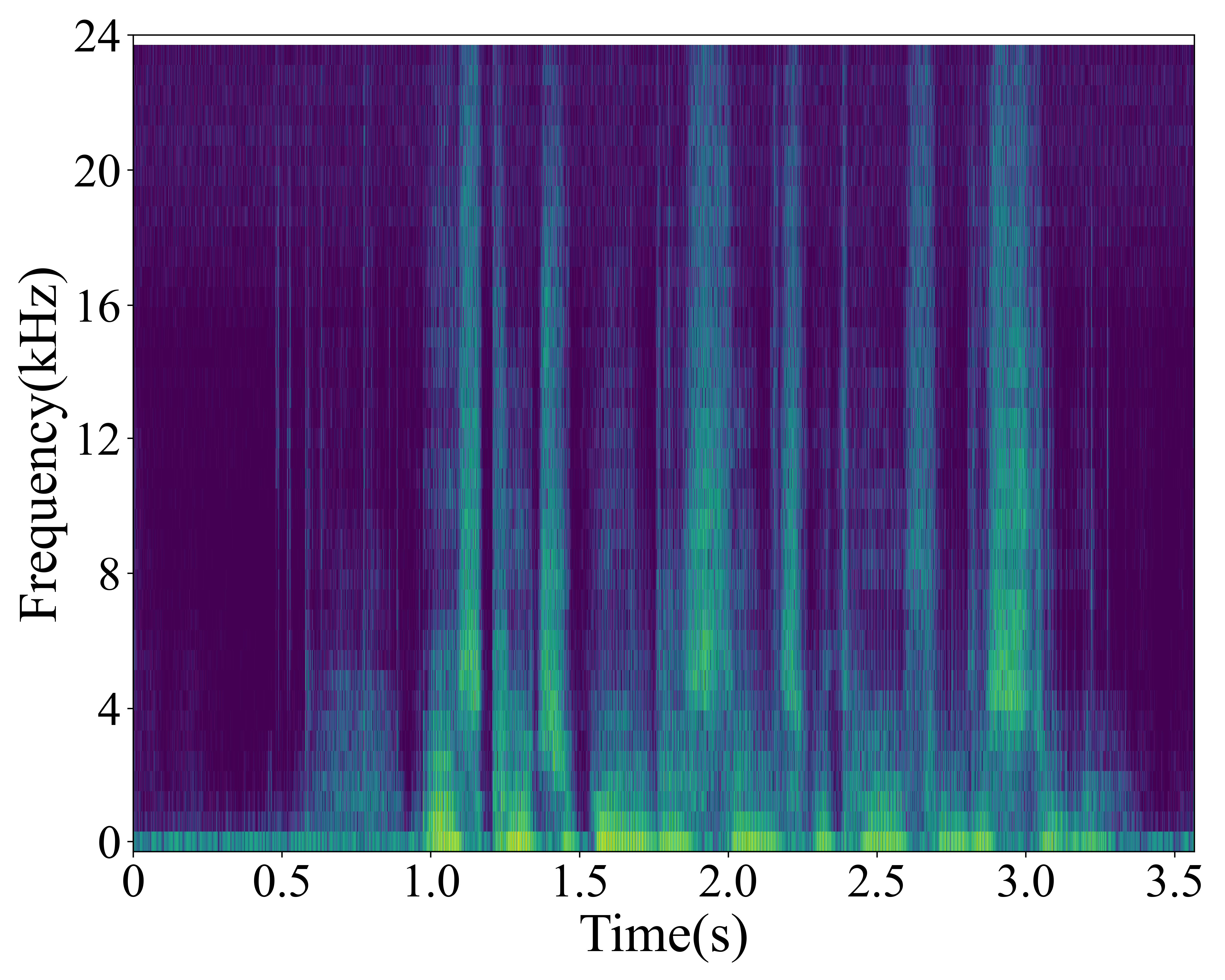}
    \label{fig:sub4}
\end{subfigure}

\caption{Visualization of natural and MDCTCodec-decoded (at 6 kbps) waveforms and MDCT spectra.}
\label{fig4}
\end{figure}
\vspace{-1mm}

\vspace{-1mm}
\subsection{Comparison with Baseline Codecs}
\vspace{-1mm}

We compared the proposed MDCTCodec with several baseline neural audio codecs, including SoundStream \cite{zeghidour2021soundstream}, Encodec \cite{defossez2022high}, HiFi-Codec \cite{yang2023hifi}, AudioDec \cite{wu2023audiodec}, DAC \cite{kumar2024high} and APCodec \cite{ai2024apcodec}. 
They were reproduced using open-source codes\footnote{\href{https://github.com/yangdongchao/AcademiCodec}{https://github.com/yangdongchao/AcademiCodec}.}\footnote{\href{https://github.com/facebookresearch/AudioDec}{https://github.com/facebookresearch/AudioDec}.}\footnote{\href{https://github.com/descriptinc/descript-audio-codec}{https://github.com/descriptinc/descript-audio-codec}.}\footnote{\href{https://github.com/yangai520/APCodec}{https://github.com/yangai520/APCodec}.}.

Experiments were conducted at a sampling rate of 48 kHz across three bitrates: 6 kbps (low bitrate), 9 kbps (medium bitrate), and 12 kbps (high bitrate). 
In addition, HiFi-Codec did not support the implementation at 9 kbps.
Therefore, it was excluded from the medium-bitrate experiment. 
MDCTNet \cite{davidson2023high} was also excluded because the range of bitrates it supported (20-32 kbps) far exceeded our configuration. 
The codecs (including MDCTCodec) all employed RVQ or related quantization strategies. 
They achieved coding at different bitrates by adjusting the number of VQs and fixing the codebook size to 1024. 




Table \ref{tab: Objective evaluation results} shows the LSD, STOI and ViSQOL scores for compared codecs. 
It is apparent that our proposed MDCTCodec exhibited 
superior performance at low and medium bitrates, particularly achieving an impressive ViSQOL score of 4.18 (the maximum score is 4.75) at 6 kbps, which denotes a significant advantage in overall decoded audio objective quality. 
At low and medium bitrates, the MDCTCodec's LSD results were suboptimal, slightly inferior to APCodec. 
One possible reason for this could be that LSD focuses on the quality of the amplitude spectrum.
APCodec directly encodes and decodes the amplitude spectrum, whereas MDCTCodec operates in the MDCT spectrum. 
Additionally, MDCTCodec significantly outperformed APCodec in terms of intelligibility, though it was slightly inferior to DAC. 
However, at higher bitrates, although MDCTCodec achieved the highest ViSQOL score of 4.32, the gap between MDCTCodec and APCodec on LSD became more pronounced, and not only DAC, HiFi-Codec also surpassed MDCTCodec in terms of intelligibility. 
This indicates that our proposed MDCTCodec is more suitable for applications in scenarios with low bitrates, possessing stronger compression capability. 
Figure \ref{fig4} also shows the visualizations of the natural and MDCTCodec-decoded waveforms and MDCT spectra at 6 kbps. 
We can see that the pattern of the MDCT spectrum and the shape of the waveform decoded by MDCTCodec are close to those of the natural ones, confirming the strong modeling capability of our proposed model.

\begin{figure}
  \centering
  \includegraphics[width=0.95\linewidth]{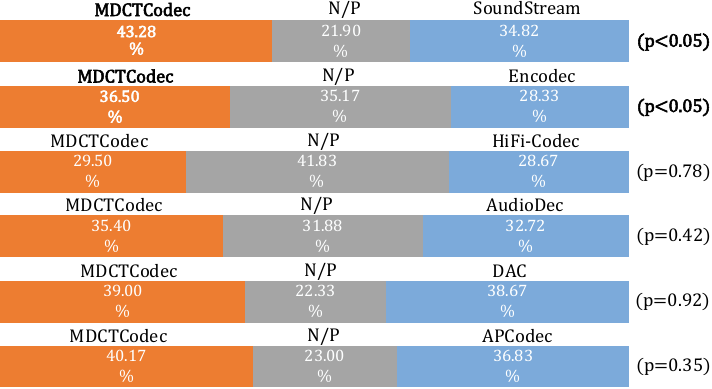}
  \caption{Average preference scores (\%) of ABX tests on speech quality between MDCTCodec and other codecs (i.e., SoundStream, Encodec, HiFi-Codec, AudioDec, DAC and APCodec), where N/P stands for "no preference" and $p$ denotes the $p$ value of a $t$-test between two codecs.}
  \label{fig: ABX}
\end{figure}

\begin{table}
  \caption{Experimental results of component analysis experiments for MDCTCodec at 6 kbps on the VCTK test set.}
  \label{tab: component}
  \centering
  \begin{tabular}{c c c c}
    \toprule
    & LSD (dB) $\downarrow$  & STOI $\uparrow$ & ViSQOL $\uparrow$ \\
    \midrule
    MDCTCodec           &       \textbf{0.825}    &0.891&\textbf{4.18}  \\
    \midrule
     w/o MDCTloss      & 0.826     &0.887&     4.13     \\
     w/o Melloss        &   1.127    &0.624&   2.84     \\
   w/o Qloss            & 2.163&0.471&  1.60    \\
    \midrule
    rep. MPD          & 0.849      &\textbf{0.912}&   3.99     \\
     rep. MRD           &0.827&0.889    &        4.16       \\
    \bottomrule
  \end{tabular}
\end{table}

\begin{table*}
\centering
    \caption{ViSQOL evaluation results for the comparison among the DAC, APCodec and MDCTCodec on the test sets of the VCTK dataset, Common Voice dataset, Opencpop dataset and FSD50K dataset at sampling rate of 48 kHz and bitrate of 6 kbps, when training from scratch on these mixed datasets. The \textbf{bold} and \underline{underline} numbers indicate optimal and sub-optimal results, respectively.}
    \resizebox{0.55\textwidth}{!}{
    \begin{tabular}{c | c c c c c c}
        \toprule
        \diagbox{Codec}{ViSQOL$\uparrow$}{Dataset} &VCTK &CommonVoice & Opencpop & FSD50K \\
        \midrule
        DAC  &3.62 &4.09&4.00&3.88 \\
        APCodec & \underline{4.10} &\underline{4.21}&\underline{4.05}&\textbf{3.95} \\
        MDCTCodec   &\textbf{4.11}  &\textbf{4.29}&\textbf{4.16}&\underline{3.93} \\
    \bottomrule
    \end{tabular}}
\label{tab_audio}
\end{table*}

Table \ref{tab: efficiency} gives results of the efficiency and complexity evaluations for compared codecs at 6 kbps. 
Our proposed MDCTCodec significantly outperformed other neural codecs in terms of generation efficiency on both GPU and CPU, as well as training time on GPU. 
Specifically, it achieved 123$\times$ and 16.9$\times$ real-time generation time on GPU and CPU, respectively. 
Particularly on the CPU, MDCTCodec's generation speed is approximately 42$\times$ faster than the slowest DAC. 
This highlights MDCTCodec's efficiency advantage of modeling spectral features over directly modeling waveform (e.g., SoundStream, Encodec, HiFi-Codec, DAC, and AudioDec) without GPU parallel acceleration. 
Compared to APCodec, MDCTCodec only needs to model the MDCT spectrum without modeling amplitude and phase spectrum in parallel, making it more efficient and boosting generation speed by approximately 2$\times$ on CPU.
The model size of MDCTCodec was also the smallest. 
This indicates that MDCTCodec is a lightweight model, allowing for its application in portable mobile devices or chips. 
Compared to the APCodec, which requires two streams for modeling both amplitude and phase, the MDCTCodec only needs one stream, thus saving approximately 2.5$\times$ the model complexity. 
Finally, the MDCTCodec possessed an extremely fast training speed, being 41$\times$ and 25$\times$ faster than HiFi-Codec and DAC, respectively. 
The total training time for MDCTCodec was only 14 hours on the VCTK dataset. 

The results of the subjective ABX tests are shown in Figure \ref{fig: ABX}.  
The MDCTCodec significantly outperformed the SoundStream and Encodec ($p > 0.05$), but there is no significant difference compared to AudioDec, HiFi-Codec, DAC and APCodec ($p < 0.05$). 
This indicates that our proposed MDCTCodec is comparable to some well-known codecs in terms of the subjective quality of decoded audio, despite having a smaller model size and faster training and generation speeds. Therefore, overall, our proposed MDCTCodec demonstrates superior performance.



\vspace{-1mm}

\vspace{-2mm}
\subsection{Component Analysis}
\vspace{-1mm}
We conducted several experiments to analyze the role of components in the proposed MDCTCodec. 
For simplicity, the experiments were conducted only at a low bitrate (i.e., 6 kbps). 
First, three ablation experiments were carried out to analyze the effectiveness of MDCT spectrum loss (i.e., $\mathcal L_{MDCT}$), mel spectrogram loss (i.e., $\mathcal L_{Mel}$) and quantization loss (i.e., $\mathcal L_{cb}$ and $\mathcal L_{com}$). 
The experimental results are presented in Table \ref{tab: component}. 
The MDCT spectrum loss only had a slight positive effect (i.e., w/o MDCTloss). 
This might be because this loss is on the order of $10^{-3}$, representing a relatively small proportion. 
Ablating mel spectrogram loss (i.e., w/o Melloss) and quantization loss (i.e., w/o Qloss) led to a significant decrease in all metrics, indicating that these losses commonly used by the majority of codecs played a fundamental role in maintaining performance. 


Furthermore, we attempted to replace MR-MDCTD with either multi-period discriminator (MPD) \cite{kong2020hifi} or multi-resolution discriminator (MRD) \cite{jang2021univnet} to validate the effectiveness of the proposed MR-MDCTD.
As shown in Table \ref{tab: component}, compared to MR-MDCTD, MPD (i.e., rep. MPD) showed an improvement in intelligibility, but there was a noticeable decline in overall audio quality. 
This suggests that the direct waveform discrimination approach of MPD is not suitable for MDCTCodec, which is a spectral-based model. 
The MR-MDCTD also outperformed MRD (i.e., rep. MRD) sightly, which may be attributed to MRD discriminating on amplitude spectra instead of MDCT spectra. 
It's worth noting that the MR-MDCTD had a fast training speed, being 3.6 times faster than MPD and 1.8 times faster than MRD.
\vspace{-1mm}
\vspace{-1mm}
\subsection{Validation of Generalization}
\vspace{-1mm}
To validate the generalization ability of MDCTCodec across different data types,
following the experiment data settings in DAC \cite{kumar2024high}, we trained DAC, APCodec and our proposed MDCTCodec by a mixed large-scale dataset with duration of approximately 1,051 hours. 
All codec models were trained from scratch on a single NVIDIA RTX TITAN GPU for approximately 120 hours.

This mixed large-scale dataset includes a wide variety of data types, specifically composed of the following four datasets. 
1) The original VCTK corpus \cite{veaux2017superseded}, still using the data split configuration mentioned in Section \ref{subsec: setting}.
2) The Common Voice dataset \cite{ardila2020common}, a large-scale multilingual transcribed speech corpus that takes approximately 919 hours. 
Compared to VCTK, its data is not as clean. 
In the experiment, we used the "Common Voice Corpus 17.0" dataset and selected all 48 kHz speeches which were then divided into 568,822 and 6,026 utterances for training and testing, respectively. 
3) The Opencpop dataset \cite{wang2022opencpop}, a publicly available high-quality Mandarin singing corpus with a sampling rate of 44.1 kHz that takes approximately 5.2 hours. 
We upsampled these data to 48 kHz for our experiments. 
We utilized the officially pre-trimmed data, selecting 3,367 utterances as the training set and the remaining 389 utterances as the test set. 
4) The FSD50K dataset \cite{fonseca2021fsd50k}, a publicly available human-labeled sound event dataset with a sampling rate of 44.1 kHz that takes approximately 84 hours. 
We upsampled these data to 48 kHz for our experiments. 
Regarding the set partition, 40,966 utterances and 10,231 utterances were respectively chosen as the training set and test set.


Although the model training is conducted on the mixed dataset, the evaluation is performed on the respective test sets of the four datasets. 
The experimental results are shown in Table \ref{tab_audio}, and only the ViSQOL scores were calculated. 
The MDCTCodec achieved the highest ViSQOL scores on the three human-vocalized datasets (i.e., VCTK, Common Voice and Opencpop). 
On the non-human vocalized dataset (i.e., FSD50K), MDCTCodec performed slightly worse than APCodec. 
This indicates that MDCTCodec is more suitable for applications related to human speech, such as speech large models, which will also be our focus in the future. 
The above experimental results confirmed the generalization capability of MDCTCodec on other types of data.

\vspace{-1mm}
\section{Conclusions}
\label{sec: Conclusion}
\vspace{-1mm}
\vspace{-1mm}
In this paper, we propose MDCTCodec, a neural audio codec which employing the MDCT spectrum as the object for encoding, quantization, and decoding. 
A novel discriminator, MR-MDCTD, is employed for adversarial training to discriminate the MDCT spectrum at various resolutions. 
Experimental results confirmed that the MDCTCodec is suitable for high-quality audio coding at high sampling rates and low bitrate scenarios. 
Additionally, it is also a lightweight model, exhibiting high training and generation efficiency. 
Further reducing the latency and applying it to downstream tasks will be the focus of our future work.

\bibliographystyle{IEEEbib}
\bibliography{strings,refs}

\end{document}